\documentclass[]{aa} 
\usepackage{graphicx}
\usepackage{natbib}
\usepackage{txfonts}
\begin{document}
\headnote{Letter to the Editor}
   \title{Detectability of dirty dust grains\\ in brown dwarf atmospheres}

   \author{Ch. Helling\inst{1}
          \and
          W.-F. Thi\inst{1,3}
          \and
         P. Woitke\inst{2}
          \and
          M. Fridlund\inst{1}
          }

   \offprints{Ch. Helling (chelling@esa.int)}

   \institute{Research and Scientific Support Department, ESTEC/ESA, 
           P.O.~Box 299, 2200 AG Noordwijk, The Netherlands\\
              \email{chelling@esa.int}
            \and
           Sterrewacht Leiden, P.O.~Box 9513, 2300 RA Leiden, The Netherlands
            \and
           Institute for Astronomy, University of Edinburgh, Blackford Hill, Edinburgh EH9 3HJ, UK}

   \date{February 1, 2006; March 11, 2006}

\abstract
{Dust clouds influence the atmospheric structure of brown dwarfs, and they affect the heat transfer and  change the gas-phase
chemistry. However, the physics of their formation and evolution is not well understood. The dust composition can be predicted
from thermodynamical equilibrium or time-dependent chemistry that takes into account seed particle formation, grain growth, evaporation and drift.} 
{In this {\em letter}, we predict dust signatures and propose a potential  observational test of the physics of dust formation in brown dwarf atmosphere based on the spectral features of the different solid components predicted by dust formation theory.}
{A momentum method for the formation of dirty dust grains (nucleation, growth, evaporation, drift) is used in application to a
static brown
dwarf atmosphere structure to compute the dust grain properties, in particular the heterogeneous grain composition and the grain size.
 Effective medium and Mie theory are used to compute the extinction of these spherical grains.}
{Dust formation results in grains whose composition differs from that of grains formed at equilibrium. Our kinetic model
predicts that solid amorphous SiO$_2$${\rm [s]}$ (silica) is one of the most abundant solid component followed by amorphous Mg$_2$SiO$_4$${\rm
[s]}$ and MgSiO$_3$[s], while SiO$_2$${\rm [s]}$ is absent in equilibrium models because it is a
metastable solid. Solid amorphous SiO$_2$${\rm [s]}$ possesses a strong broad absorption feature centered at 8.7~$\mu$m, while
amorphous Mg$_2$SiO$_4$${\rm [s]}$/MgSiO$_3$[s] absorb at 9.7~$\mu$m beside other absorption features at longer wavelength. Those features at
$\lambda < 15\mu$m are detectable in absorption if grains are small (radius $<$~0.2~$\mu$m) in the upper atmosphere as suggested
by our model.}
{We suggest that the detection of a feature at 8.7~$\mu$m in deep infrared spectra could provide evidence for non-equilibrium dust
formation that yields grains composed of metastable solids in brown dwarf atmospheres. This feature will shift towards
$10\mu$m and broaden if silicates (e.g. fosterite) are much more abundant.}
 
   \keywords{brown dwarfs - spectral features}

   \maketitle
%

\section{Introduction}~\label{introduction} 

\noindent
Atmospheric models are essential in the interpretation of complex brown dwarfs spectra, which are dominated by strong molecular
absorption lines. 
The physics and chemistry of substellar objects (brown dwarf and giant planets) are more complicated than anticipated because of
non-equilibrium processes (dust formation, convective mixing) and their non-linear feedbacks on the radiative
transfer. Consequently, parametrisations of processes like dust formation and convection were applied in atmospheric models so far.
Current model atmospheres provide extensive solutions of the radiative transfer problem for the gas phase in hydrostatic equilibrium \citep{Tsuji1996A&A...308L..29T,Marley1996Sci...272.1919M,Burrows1997ApJ...491..856B,Allard2001ApJ...556..357A}.
However, the presence of dust is at present only inferred from classical models, where the dust composition is
derived from gas/solid phase equilibrium considerations and time-scale arguments, and their comparison to  observed spectra
\citep{Lunine1989ApJ...338..314L,Tsuji1996A&A...308L..29T,Tsuji2002ApJ...575..264T,Ackerman2001ApJ...556..872A,Allard2001ApJ...556..357A,Cooper2003ApJ...586.1320C}.
Such equilibrium chemistry results in grains with a homogeneous composition.

Considering the non-equilibrium character of phase transitions (supersaturation $\!\!\gg\!\!1$), \cite{Woitke2003A&A...399..297W,Woitke2004A&A...414..335W} and
\cite{Helling2005} proposed a theoretical approach to consistently model the formation of dust grains by nucleation (seed
formation), growth, evaporation and drift (gravitational settling).  In contrast to the phase-equilibrium calculation, Woitke \&
Helling argue that grains in the oxygen-rich environment of a brown dwarf are heterogeneous in chemical composition and in
size. Moreover, precipitating into the denser inner atmospheres, these particles can reach sizes of several $100$~$\mu$m. In
Helling \& Woitke (2006), the inferred dust composition differs markedly from that predicted by equilibrium chemistry. These
differences manifest in the intrinsic absorption signatures of the solids in the mid-infrared range.

This {\em letter} suggests a potential observational test for the presence of dust. The dust may be present in form of cloud-like
structures. We will focus on the main characteristics of the theory, where the predictions in terms of dust spectral features can
be tested with current and future instruments. In Sect.~\ref{approach}, we report on recent progress in modeling the formation of
chemically non-homogeneous cloud layers made of dirty (i.e. a variety of compounds) grains in brown dwarf atmospheres. A detailed Mie theory treatment combined
with effective medium theory allows us to suggest possible spectral features of such a dust layer for both approaches. The result
of the modeling is described and discussed in Sect.~\ref{results}

\section{Approach}~\label{approach} 

\vspace*{-0.3cm}
\noindent We model nucleation, heterogeneous growth, evaporation, and drift (gravitational settling) of dirty dust particles in a
quasi-static atmosphere by the moment method \citep{Helling2005}. Our prescription considers the formation of compact 
spherical grains in an oxygen-rich gas by the initial nucleation of TiO$_2$ seed particles followed by the  growth of
a dirty mantle made of various solids. We consider amorphous TiO$_2$${\rm [s]}$, SiO$_2$${\rm [s]}$,
MgO[s], MgSiO$_3$[s], Mg$_2$SiO$_4$${\rm [s]}$, Al$_2$O$_3$${\rm [s]}$, and metallic iron Fe${\rm [s]}$, assuming that silica and
silicates have approximately the same sticking probability. The moment and elemental conservation equations
for Ti, Si, Mg, Fe, Al, and O are evaluated on top of a prescribed static model atmosphere structure (Allard et
al.~\footnote{adopted from ftp.ens-lyon.fr/pub/users/CRAL/fallard/}) without further iteration on the temperature profile.  The model calculates the amount of condensates, the mean grain size
$\langle a \rangle$, and the volume fractions $V_{\rm s}$ of each material as function of height $z$ in the brown dwarf atmosphere
for given model structures.  At each depth in the atmosphere, the volumes fractions are used to compute mean dust optical
constants using the Maxwell-Garnett effective medium theory \citep{Bohren1983asls.book.....B}, which is a valid method because the
grains generally have one main volume component. 
The main dust component is the matrix in which the spherical
inclusions made of the minor components are embedded. Dust extinction coefficients are subsequently computed using a Mie theory
code for compact spherical grains.  The different components are assumed in an amorphous state, consistent with laboratory
experiments of heterogeneous dust formation (e.g., \citealt{Rietmeijer1999ApJ...527..395R}).  The optical constants for the
amorphous solids are obtained from laboratory measurements (TiO$_2$${\rm [s]}$: \cite{Ribarsky85}, SiO$2$$_{\rm [s]}$:
\cite{Henning1997A&A...327..743H}, MgO: \cite{Hofmeister2003MNRAS.345...16H}, MgSiO$_3$[s]/Mg$_2$SiO$_4$${\rm [s]}$: \cite{Jager2003A&A...408..193J}, Al$_2$O$_3$${\rm [s]}$:
\cite{Begemann1997ApJ...476..199B}, and Fe${\rm [s]}$: \cite{Ordal1985ApOpt..24.4493O}).  Absorption cross sections and optical
depth are then derived using the dust density profiles assuming a grain size distribution $f(a,z)=\langle a(z)
\rangle\,\delta(a-\langle a(z) \rangle)$. For comparison, similar computations were performed assuming that no SiO$_2{\rm [s]}$
can form by setting the formation rate to zero. We also estimate the possible absorption depth in observed spectra. The contrast
between the depth of the features and the continuum is an important parameter for predicting the detectability of the dust
features.

\section{Results}~\label{results}

\noindent 
We first show the grain size and dust composition profile of a AMES (cond) model with T$_{\rm eff}=1500\,$K, $\log$ g$=5.0$
($^1$) and solar element composition. This can be considered a typical
late L-dwarf atmosphere model. The outputs are then used to compute the dust extinction coefficient in the mid-infrared.

\subsection{Global grain size distribution and material composition} 

\begin{figure*}[!ht]
   \centering
   \includegraphics[width=8.6cm, height=7cm]{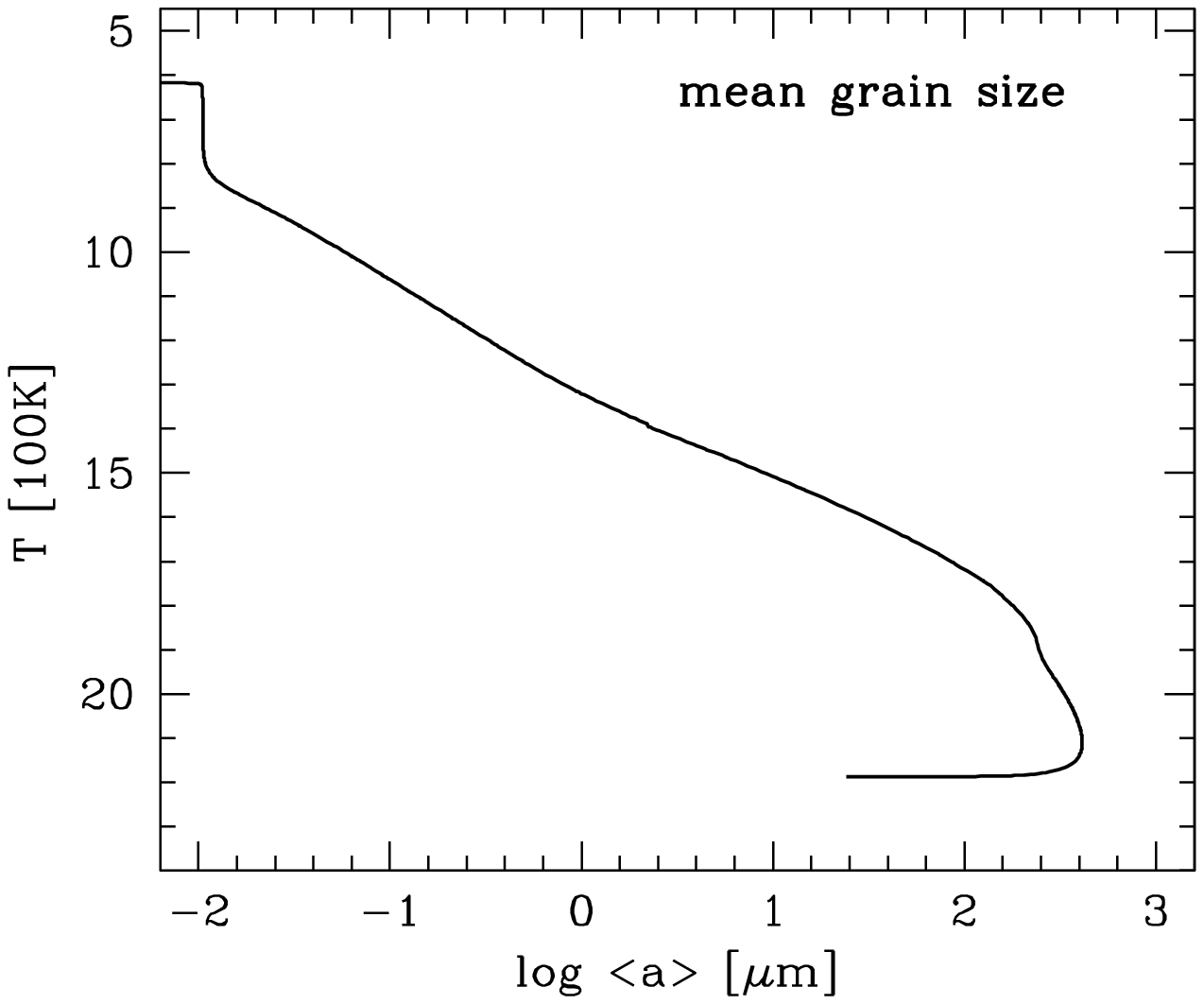}
   \includegraphics[width=8.6cm, height=7cm]{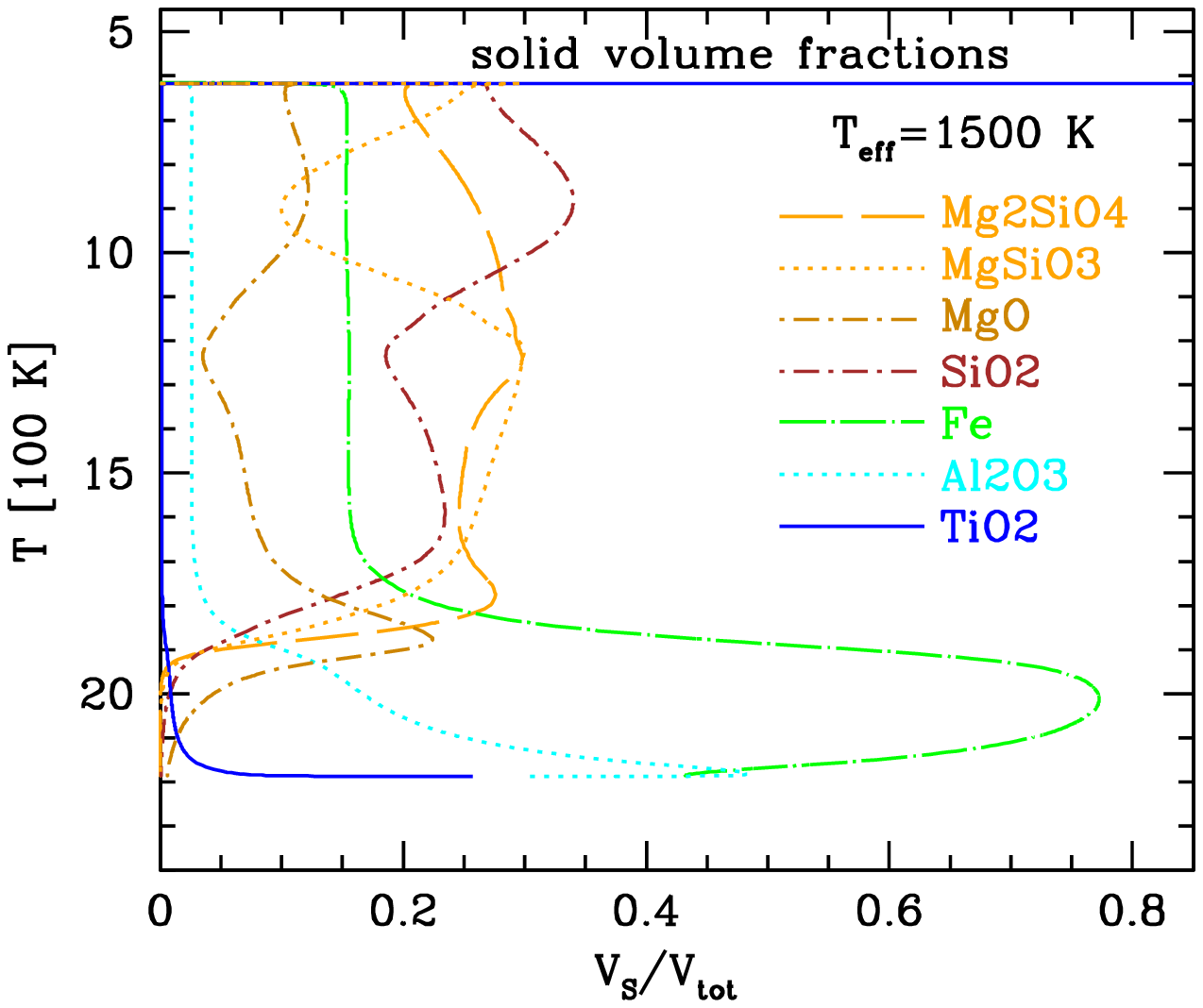}
   \vspace*{-0.1cm}
      \caption{{\bf Left:} Mean grain size  stratification for a 
    plan-parallel AMES (cond) late L-type brown dwarf model atmosphere of T$_{\rm eff}=1500\,$K, $\log$\,g$=5.0$ of original solar
    abundance. {\bf Right:} Chemical material composition of the grains in volume fractions of
   the solids for the same model ($V_{\rm tot}^{\rm dust}=\sum_{\rm s} V_{\rm s}$, s -- contributing solid materials). Note the upside-down scale of the local temperature T [100\,K].
              }
  \label{meanSize}
   \vspace*{-0.3cm}
   \end{figure*}

The results depicted in Fig.~\ref{meanSize} show the same global dust cloud structure as presented in
\cite{Woitke2004A&A...414..335W}. The formation of the cloud is governed by the hierarchical dominance of nucleation (upper most
layers), growth \& drift (intermediate layers), and evaporation \& drift (deepest layers).  The uppermost cloud layers are
predominantly filled with small grains of a mean size of $10^{-2}\mu$m which grow on their way into the atmosphere to 
$200\,\dots 300\,\mu$m (l.h.s. Fig.~\ref{meanSize}). Eventually they enter even hotter atmospheric layers where they are no
longer thermally stable.  Hence, the grains shrink in size and finally dissolve into the surrounding hot and convective gas. This
picture reflects the stationary character of the grain component forming the brown dwarf's dust cloud where dusty material
constantly fall inward and fresh, dust-free material is mixed upward.

\begin{table}[!ht]
\begin{center}
\caption{Dust grain's mean size and material composition.}
~\label{table_summary}
\vspace*{-0.4cm}
\begin{tabular}{lll}
\hline
\hline
\noalign{\smallskip}
local             & mean grain & material\\
temperature  & size  & composition\\
T [K] & $\langle a\rangle$ [$\mu$m] & [volume frac.]\\
\noalign{\smallskip}
\hline
\noalign{\smallskip}
$\lesssim 1000$& $10^{-2}\,\ldots\,10^{-1}$ &  $25 - 35\%$ SiO$_2$[s]\\
  & & $20 - 25\%$ Mg$_2$SiO$_4$[s]\\
  & & $10 - 25\%$ MgSiO$_3$[s]\\
  & & $15\%$ Fe[s], $10\%$ MgO[s]\\[0.1cm]
$1000\,\ldots\, 1700$& $10^{-1}\,\ldots\,30$ & $25 - 30\%$ Mg$_2$SiO$_4$[s]\\
  & & $20 - 30\%$ MgSiO$_3$[s] \& SiO$_2$[s]\\
  & & $15\%$ Fe[s],  $10\%$ MgO[s]\\[0.1cm]
$1700\,\ldots\,2000$ & $30\,\ldots\,200$ & strongly changing composition\\[0.1cm]
$2000\,\ldots\,2300$ & $200\,\ldots\,300$ & $80\%$Fe[s], $\sim 18\%$ Al$_2$O$_3$\\
\noalign{\smallskip}
 \hline
\end{tabular}
\end{center}
\end{table}
   \vspace*{-0.3cm}

The r.h.s. of Fig.~\ref{meanSize} demonstrates the chemical composition of the dust grains.  The grains are not of a single,
homogeneous material composition and their material composition changes on their descent into the atmosphere.  Since gas and dust
thermalise faster than the dust growth/evaporation processes take place (\citealt{Woitke2003A&A...399..297W}), the chemical
composition is determined by the local temperature and the reaction kinetics.  Therefore, the upper atmosphere is populated by
small, silica- and silicate-like grains until these materials become thermally unstable. Therefore in the lower and hotter part of the
atmosphere big grains appear which are merely made of iron and some impurities of Al$_2$O$_3$${\rm [s]}$ and TiO$_2$${\rm
[s]}$. The main dust component in the upper part of the dust layer is amorphous SiO$_2$${\rm [s]}$ followed by Mg$_2$SiO$_4$${\rm
[s]}$/MgSiO$_3$[s] (see Table~\ref{table_summary}).  The prediction contrasts with those from equilibrium dust formation
models where metastable species such as SiO$_2$${\rm [s]}$ cannot exist (e.g Lodders \& Fegley 2005). 

\subsection{Spectral dust cloud features} 

\begin{figure}[!ht]
   \vspace*{-0.4cm}
   \hspace*{-0.3cm}
   \includegraphics[width=7.6cm, height=9.4cm,angle=90]{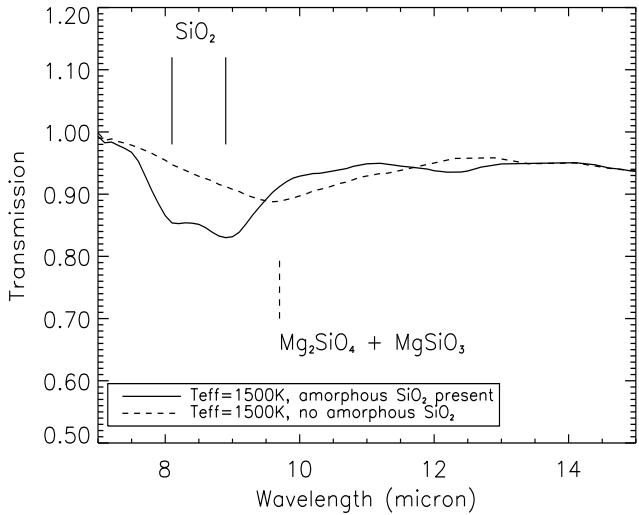}\\[-0.7cm]
   \caption{Absorption transmission spectrum ($=F_{\nu}/F_{\rm cont}$ with $F_{\rm cont}$ the local continuum flux with respect to the
   feature) for the model in Fig.~\ref{meanSize} with (solid line) and without (dashed line) formation of SiO$_2$${\rm
   [s]}$. The absorption feature of
   SiO$_2$${\rm [s]}$ centered at $\sim$8.7 $\mu$m.
         \label{Spec}}
   \vspace*{-0.3cm}
   \end{figure}

Figure~\ref{Spec} depicts the resulting spectral features from   6 $\mu$m $\ldots$ 15
 $\mu$m for a brown dwarf dust cloud layer with the mean grain size and chemical dust material composition shown in Fig.~\ref{meanSize}.

For grain composition predicted by heterogeneous dust formation, the only dust features with an appreciable contrast are those of
SiO$_2$${\rm [s]}$ centered at 8.7~$\mu$m and of Mg$_2$SiO$_4$${\rm [s]}$/MgSiO$_3$[s] at 9.7~$\mu$m, with weaker absorption
features around 20~$\mu$m and ~32~$\mu$m (not shown). The features are broad ($\sim$~1 -- 1.5 ~$\mu$m) and lack substructures
because the grains are amorphous. The abundance of metallic iron is high ($\sim$~15\%) but metals absorb photons continuously and
do not show spectral features. The abundance of Al$_2$O$_3$${\rm [s]}$ and TiO$_2$${\rm [s]}$ in grains are too small to
significantly affect the overall extinction coefficient. In models where SiO$_2$${\rm [s]}$ is disregarded (dashed line,
l.h.s. Fig.~\ref{Spec}), the dust extinction is dominated by Mg$_2$SiO$_4$${\rm [s]}$/MgSiO$_3$[s] at 9.7~$\mu$m.  The mean grain radius
remains smaller than 0.2~$\mu$m in the $\tau_{\rm dust}<$~1 region in the model. Therefore, the grain absorption features
remain. In contrast, if grains were a large as $10~\mu$m, the resonance features would disappear \citep{Min2004A&A...413L..35M}.
The region where $\tau_{\rm dust}<$~1 in the wavelength range 5 -- 25 $\mu$m has pressure below 10$^{-2}$ bar in the model
atmosphere considered.

In phase-equilibrium, the condensates are formed locally and retain the energetically most favorable composition at the
temperature and pressure of the given depth in the atmosphere \citep{Lodders2005}. Further condensation is not possible because
the grains remain in relative thin discrete cloud layers and, hence,  can not move into regions with still favorable
growth conditions. Individual cloud layers are composed of different solid species
depending on the temperature of the layer. 
Above the silicate clouds,  the gas phase is strongly depleted and thus
the absence of gas-phase species in L and T dwarf atmosphere has been used as an evidence for the presence of clouds. 
The
prediction of equilibrium theory is the absence of solid SiO$_2$, hence equilibrium-dust is predominately composed of
silicates. We modeled the absence of SiO$_2$${\rm [s]}$ and the spectral result is shown in Fig.~\ref{Spec} (l.h.s.). Beside moving
the peak of the absorption to 9.7~$\mu$m, the absence of amorphous SiO$_2$$_{\rm [s]}$ decreases substantially the absorption
contrast.

\subsection{Discussion of detectability}~\label{detectability}

\vspace*{-0.4cm}
\noindent We now discuss the detectability of infrared dust features in brown dwarf atmospheres. From Fig.~\ref{Spec} (r.h.s.), we
can see that the absorption over continuum contrast varies from 10$\,\ldots\,$ 20\% with the maximum obtained for cold brown
dwarfs if SiO$_2$${\rm [s]}$ is present. This contrast is achieved because the grains are small ($<$~0.2~$\mu$m). In the absence
of SiO$_2$${\rm [s]}$, Mg$_2$SiO$_4$${\rm [s]}$/MgSiO$_3$[s] will absorb weakly at 9.7~$\mu$m. In the case of large grains, no
feature can be detected and it will not be possible to determine the composition of these grains. The presence or absence of
SiO$_2$${\rm [s]}$ can be tested observationally. Recently, \cite{Cushing2006ApJ} presented {\it Spitzer}
IRS\footnote{$R=90$ for $\lambda = 5.3\mu$m$\,\ldots\,15.3~\mu$m} results showing L-dwarf spectra with a $1 - 1.5~\mu$m broad
absorption centered at
$\lambda\approx 9~\mu$m.  However, we have treated an
idealized situation where dust grains are the main sources of opacity. In reality, the feature will appear on the top of
gas absorption lines.  
The gas phase absorption proceeds by
lines whose width is determined by the pressure at the height of the absorption layer. Typical molecular band widths are
$\sim$~0.1~$\mu$m.  

Another limitation of our treatment is the assumption of spherical grains which may introduce spurious effects during
the opacity calculation.  Grains are most likely non-spherical and their absorption coefficients should be computed with, for
example, the hollow sphere model \citep{Min2005A&A...432..909M} which successfully models asphericity by applying a distribution
of hollow spheres. However, the use of hollow spheres will introduce minor changes in the shape of the absorption feature and will
not change significantly the wavelength-position of the dust features and hence the conclusion of this paper.  Finally, the
heterogeneous formation model does not compute the exact grain size distribution but only its moments from which the mean size can
be computed. Future work is needed to calculate better analytical representations of the size distribution function from the dust
moments.

\section{Conclusions}~\label{conclusions}

\vspace*{-0.3cm}
\noindent 
We have studied the dust spectral signatures predicted by the Helling \& Woitke (2006) models which make specific observational
predictions in terms of grain composition and size at different altitudes.  Heterogeneous dust formation in brown dwarf atmosphere
results in dusty grains composed mainly of amorphous SiO$_2$${\rm [s]}$ and amorphous Mg$_2$SiO$_4$${\rm [s]}$/MgSiO$_3$[s] with
impurities of Fe and Al oxides, TiO$_2$ and MgO. The dust forms one continuous layer in which the grains gradually change
composition and size. The grains are sufficiently small in the upper atmosphere to produce solid resonance features in the
infrared, a conclusion reached also by  \cite{Cushing2006ApJ}. In contrast, grain composition derived from chemical equilibrium models is dominated by Mg$_2$SiO$_4$${\rm [s]}$ with
the noticeable absence of amorphous SiO$_2$${\rm [s]}$, which is a metastable species.  The difference in dust composition allows
us to make predictions in dust spectral features in the infrared for the two types of grains. Amorphous SiO$_2$${\rm [s]}$ moves
the peak of the absorption to 8.7$\mu$m while a nearly pure Mg$_2$SiO$_4$${\rm [s]}$/MgSiO$_3$[s] absorbs at 9.7$\mu$m with 
a much smaller contrast. Therefore,
infrared features of dust grains could be used to directly detect the presence of dust  in
brown dwarf atmospheres.  The detection of dust absorption features supports the prediction that grains have small size
$<$~0.5~$\mu$m and is present in the $\tau_{\rm dust}<$~1 upper atmosphere.

\begin{acknowledgements} We thank the anonymous referee for her/his suggestion. We thank A.~Heras for commends on the manuscript,
F.~Allard and P.~Hauschildt for making their model results freely available. ChH and WFT are/were supported by ESA internal fellowships
at ESTEC. PW was supported by the NWO Computational Physics program, grant 614.031.017, in the frame of the {\sc AstroHydro3D}
initiative. The computer support by the ESTEC RSSD computer team is highly appreciated.  Most of the literature search was done
with ADS.  \end{acknowledgements}


\end{document}